\documentclass[showpacs,preprintnumbers,amsmath,twocolumn,amssymb,revsymb,prl, superscriptaddress,aps]{revtex4-1}
\usepackage{graphicx}
\usepackage{amsmath}
\usepackage{dcolumn}

\usepackage{float}
\usepackage{dcolumn}
\usepackage{bm}%
\usepackage{mathptmx}
\usepackage{titlesec}
\usepackage{xcolor}
\usepackage{lipsum}

\begin{document}
\title{ Parallel Quantum Circuit in a Tunnel Junction}

\author{Omid Faizy Namarvar } 
\affiliation{ CEMES-CNRS, 29 rue J. Marvig, 31055 Toulouse Cedex,
France
}

\author{Ghassen Dridi }  
\affiliation{ CEMES-CNRS, 29 rue J. Marvig, 31055 Toulouse Cedex,
France
}

\author{Christian Joachim }  
\affiliation{ CEMES-CNRS, 29 rue J. Marvig, 31055 Toulouse Cedex,
France
}
\affiliation{ WPI-MANA, National Institute for Material Sciences, 1-1 Namiki, Tsukuba, Ibaraki, Japan
}

\
\date{\today{}}

\begin{abstract}

The spectrum of 1-state and 2-states per line quantum buses is used to determine the effective $V_{ab}(N)$ electronic coupling between emitter and receiver states through the bus as a function of the number $N$ of parallel lines in the bus. When the calculation of $V_{ab}(N)$ is spectrally difficult, an Heisenberg-Rabi time dependent quantum exchange process can be triggered through the bus by preparing a specific initial non-stationary state and identifying a target state to capture the effective oscillation frequency $\Omega_{ab}(N)$ between those. For $\Omega_{ab}(N)$ (for $V_{ab}(N)$), two different regimes are observed as a function of $N$:  linear and  $\sqrt{N}$ more moderate increases. This state preparation was replaced by electronically coupling the quantum bus to two semi-infinite electrodes. The native quantum transduction process at work in this tunnel junction is not faithfully following the $\Omega_{ab}(N)$ variations with $N$. Due to normalisation to unity of the electronic transparency of the quantum bus and to the low pass filter character of the transduction, large $\Omega_{ab}(N)$ cannot be followed by the tunnel junction. At low coupling and when $N$ is small enough not to compensate the small through line coupling, an $N^2$ power law is preserved for $\Omega_{ab}(N)$. The limitations of the quantum transduction in a tunnel junction is pointing how the broadly used concept of electrical contact between a metallic nanopad and a molecular wire can be better described as a quantum transduction process. 
\end{abstract}



\maketitle


\begin{section}{1) Introduction}

Installing a quantum transfer line in between two identical $A $ and $B$ quantum systems opens the possibility to transfer from $A$ to $B$ one electron added to $A$ because of the electronic coupling introduced between $A$ and $B$ by this line \cite{joachim87,chen}. To increase the chance for this electron to be transferred, to speed up its transfer or to minimize the energy required, more transfer lines can be added in parallel forming a quantum bus between A and B \cite{chen}. In absence of electronic coupling between the lines, $N$ identical lines in parallel must intuitively increase the $V_{ab}(N)$  coupling between state $\vert\phi_a\rangle$ (electron on $A$) and state $\vert\phi_b\rangle$ (the electron on $B$). Quantifying the $V_{ab}(N)$ power law  of this superposition and measuring it experimentally are long standing problems \cite{joachim87,chen}. A possible measure (i) is to perform a spectroscopy characterization of the $A-N-B$ quantum system to follow how the free from bus $\vert\phi_a\rangle$, $\vert\phi_b\rangle$ degeneracy is lifted up by the progressive insertion of $N$ lines in parallel between $A$ and $B$. Measure (ii) protocol is to follow in real time the electron transfer process between $A $ and $B$ and to measure how $N$ is changing the $\Omega_{ab}(N)$ Heisenberg-Rabi secular oscillation frequency of this process before any relaxation (for example the electron being trapped on $A$ (on $B$) or ejected from $A-N-B$). Measure (iii) protocol is to connect $A$ and $B$ to metallic nanopads $M_{A}$ and $M_{B}$ interacting respectively with $\vert\phi_a\rangle$ and $\vert\phi_b\rangle$, to low bias voltage the corresponding $M_{A}$-A-N-B-$M_{B}$ junction and to follow the variations of the $I(N)$ current intensity through this junction as a function of $N$.

Measure (i) had long been practiced since the first electron transfer experiments through molecular wires \cite{launay} and had more recently been used for example in mesoscopic qubit systems \cite{nakamura} and to measure the electronic coupling between 2 metallic nano-cubes stabilized together by a small number $N$ of short molecular wires self-assembled in parallel \cite{tan}. For large $V_{ab}(N)$, (i) has the inconvenience that recovering the $\vert\phi_a\rangle$ and $\vert\phi_b\rangle$ states in the complete $A-N-B $ electronic spectrum is quite difficult because in this case, $\vert\phi_a\rangle$ and $\vert\phi_b\rangle$ are very much diluted over the $A-N-B$ eigenstates. 

Measure (ii) is depending on the technical possibility to follow in real time very fast phenomena since even for $V_{ab}(N)$ of the order of a few $\mu eV$, $\Omega_{ab}(N) = 2V_{ab}(N)/ \hbar$ \cite{joachim_CF_87} can already reach the GHz regime \cite{nakamura,zewail}. In case of quantum decoherence along the bus (for example $\vert\phi_b\rangle$ not fully reconstructed in time on $B$ after the initial preparation of $\vert\phi_a\rangle$ on $A$), it is very difficult to sort out $\Omega_{ab}(N)$ because in this case, the time evolution of the $\vert\phi_b\rangle$ population will only be almost-periodic \cite{sautet_88}. 

Measure (iii) is intermediate between (i) and (ii) because as demonstrated in this paper, $I(N)$ is in effect the long time average (low pass filtered) transduction of the $\vert\phi_b\rangle$ time evolution population amplitude normally tracked by (ii). Furthermore, (iii) is not a static characterization of the $A-N-B$ spectrum like in protocol (i) which is looking for the $\vert\phi_a\rangle$ to $\vert\phi_b\rangle$ energy splitting among the $A-N-B$ eigenstates. 

For low $V_{ab}(N)$ and by generalization of the Bardeen perturbation approach of tunneling by including quantum states in the tunneling barrier \cite{bardeen}, it was long demonstrated that in the tunneling regime $I(N) = N^2 \, J$  where $J$ is the elementary tunneling current intensity passing through a single transfer line of the quantum bus \cite{joachim,magoga}. This simple $N^2$ power law was recently questioned because for some specific  molecular scale quantum bus, $I(N)$ was found to be even lower than elementary $J$ \cite{lambert} while in other experiments, it was proven to be valid at least for small $N$ \cite{vazquez}. 

To clarify the situation, we propose in this paper a complete demonstration and analysis of the $I(N)$ variations resulting from an increase of the electronic coupling between $A$ and $B$ as a function of $N$. Introducing the exact quantum transduction function to pass from the $\vert\phi_b\rangle$ population amplitude to the $T(E_{f},N)$ electronic transparency of the $M_{A}$-A-N-B-$M_{B}$ tunneling junction (nano pads Fermi energy $E_{f}$), we show how measurement (iii) has one drawback explaining why for large $V_{ab}(N)$, the $N^2$ power law was recently questioned. At low bias voltage and according to the Landauer formula, $I(N)$ is proportional to $T(E_{f},N)$. But $T(E_{f},N)$ is necessary bound from above to unity. As a consequence, large $V_{ab}(N)$ values cannot be measured using (iii) demonstrating how (iii) is not doing much better than (i) for tracking the power law of $V_{ab}(N)$ for large $V_{ab}(N)$ values. Two types of quantum bus are used for this demonstration, with one and two quantum states per line. In section 2, the spectral analysis of the corresponding $A-N-B$ quantum Hamiltonians and of the time dependant quantum evolution after preparing $A-N-B$ initially in the non stationary state $\vert\phi_a\rangle$ are provided in a way to determine the $V_{ab}(N)$ variations as a function of $N$ and of the bus control parameters. In Section 3, the exact transformation between the $\vert\phi_b\rangle$ time dependant population amplitude and $T(E_{f},N)$ is presented showing how this transformation is a quantum to classical low pass filter transduction between a quantum time dependant phenomenon and the tunneling junction conductance. In section 4, this transformation is used to provide the limit of validity of (iii) determining when the $N^2$ law can be applied and what is measured if not. In conclusion, the consequence of the of the limitations of the quantum transduction at work in a tunneling junction are discussed in the perspective of improving the contact conductance between a molecular wire and its metallic nanopads.

\end{section}


\begin{section}{2) Spectral analysis and time dependent Heisenberg-Rabi oscillations}

To interconnect $A$ and $B$ by a quantum bus and to be able to use analytical solutions to determine $V_{ab}(N)$, only two type of multipath quantum systems are considered in the following with 1-state and then 2-states per transfer line. A number  $N$ of those lines are interacting in parallel, equally and independently with state $\vert\phi_a\rangle$ and $\vert\phi_b\rangle$. A quantum bus with $N$ 1-state per line is the first member of a family having an odd number of states per line i.e. with always one eigenstate of the corresponding bus Hamiltonian located in the middle of its spectrum. A quantum bus with 2-states per line is the second member of a family having an even number of states per line i.e. having no state in the middle of its spectrum  \cite{lang}. The first member of this second family is simply the direct through space coupling between $A$ and $B$. For a quantum bus, having or not an eigenstate located in the middle of its electronic spectrum has profound consequences on the measurability of large $V_{ab}(N)$ values through this bus.


\begin{subsection}{2.1) $N$ transfer lines in parallel with 1-state per line}

On the $A-N-B$ canonical basis set $\vert\phi_a\rangle$, $\vert j \rangle$ ($j=1,N$) and $\vert\phi_b\rangle$, Fig. \ref{fig:model1} is presenting the complete N+2 quantum states graph of the quantum bus with N 1-state per transfer line interacting with the emitter state $\vert\phi_a\rangle$ and the receiver state $\vert\phi_b\rangle$. Each 1-state line is $\gamma$ interacting equally with $\vert\phi_a\rangle$ and $\vert\phi_b\rangle$ and there is a relative energy difference $\Delta$ between the quantum bus states and $\vert\phi_a\rangle$, $\vert\phi_b\rangle$. This defines two quantum $\gamma$, $\Delta$ and one classical $N$ control parameters for the $A-N-B$ system.

 \begin{figure}[!h]
    \centering
    \includegraphics[scale=0.25]{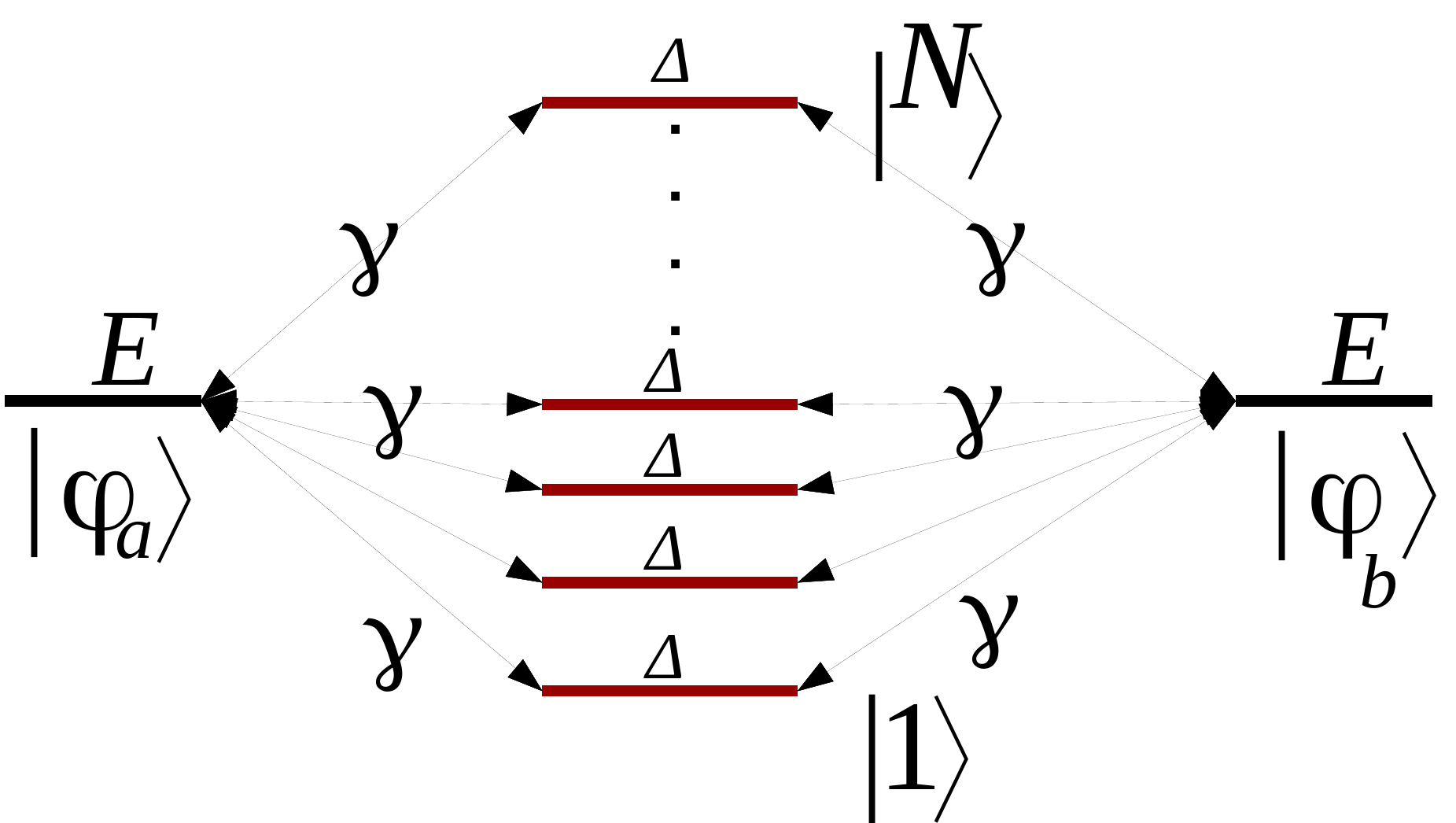} 
   \caption{The quantum graph of an $N$ 1-state per line bus interconnecting $A$ and $B$ with $\vert\phi_a\rangle$ for the electron on $A$, $\vert\phi_b\rangle$ the electron on $B$ and $\vert j\rangle$ for the electron on the bus states. This determines the valence bond like canonical basis set of the system. The $N$ parallel states have the same electronic energy $\Delta$ relative to $\vert\phi_a\rangle$ and $\vert\phi_b\rangle$ and are interacting with $\vert\phi_a\rangle$ and $\vert\phi_b\rangle$ via the electronic coupling $\gamma$.}
  \label{fig:model1}
  \end{figure}

The quantum properties of the Fig.\ref{fig:model1} system have already been studied in detail for the purpose of engineering a bistable switch after playing with the electronic coupling of one transfer line \cite{joachim87}. We recall in this section the essential characteristics of this system not for switching but to focus on another aspect of its quantum controllability: the speed up of the electron transfer between $A$ and $B$ as a function of $N$. On its canonical basis set, the mono-electronic Hamiltonian of the Fig.\ref{fig:model1} system is given by \cite{joachim87}:
  
\begin{equation} \label{eq:hamiltonian1}
H(N,\Delta,\gamma)= \left[ 
    \begin{matrix}
      0 & 0 & \gamma & \gamma &\cdots& \gamma  \\
      0 & 0 & \gamma & \gamma &\cdots& \gamma  \\
      \gamma & \gamma & \Delta & 0 &\cdots& 0\\
       \gamma & \gamma & 0 & \Delta &\cdots& 0\\
       \vdots & \vdots & 0 & 0 &\ddots& \vdots\\
        \gamma & \gamma & 0 & 0 &\cdots& \Delta
      \end{matrix}
    \right].      
\end{equation}
Its spectrum has $N+2$ eigenvalues, $N-1$ being degenerate of value $\Delta$, one $\lambda_3=0$ and the two remaining ones $\lambda_1$ and $\lambda_2$ are given by:
  \begin{equation} \label{eq:spectrum1}
  \lambda_{1,2}=\dfrac{1}{2}\left[\Delta\pm\sqrt{\Delta^2+8N\gamma^2}\right]
  \end{equation}

For $\gamma << \Delta$, only two of those eigenvalues have their corresponding eigenvector very close to $\vert\phi_a\rangle$ and $\vert\phi_b\rangle$. In this case, the effective through bus coupling $V_{ab}(N)$ is simply 1/2 the energy splitting between $\lambda_2$ and $\lambda_3$ leading to $V_{ab}(N)\simeq \dfrac{N\gamma^2}{\Delta}$ which is increasing linearly with $N$. For $\gamma > \Delta$ or for $\Delta=0$, the search for those two eigenvectors in the $H(N,\Delta,\gamma)$ spectrum is more difficult. For example for $\Delta=0$, the $\lambda_3$  corresponding eigenvector has still the highest weight on $\vert\phi_a\rangle$ and $\vert\phi_b\rangle$. But at the same time, $\lambda_1$  and $\lambda_2$ have exactly the same weight. In the intermediate regime where $\gamma$ and $\Delta$ are of the same order of magnitude, $\lambda_2$ is still the second leading one and $V_{ab}(N)=(\lambda_{2}-\lambda_{3})/2=\dfrac{1}{4}[\Delta-\sqrt{\Delta^2+8N\gamma^2}]$ i.e. an $\sqrt{N}$ law for $V_{ab}(N)$.

Following protocol (ii), one way to determine $V_{ab}(N)$ in all the $\gamma$ and $\Delta$ cases is to prepare the Fig. \ref{fig:model1} system at $t=0 $ in the non stationary state $\vert\phi_a\rangle$ to trigger a spontaneous response of the complete $A-N-B$ system in time and to determine the effective $\Omega_{ab}(N)$ oscillation frequency of the transfer process. As compared to the above spectral analysis for 
tracking $V_{ab}(N)$, the advantage of this preparation is that $\vert\phi_a\rangle$ is now specified and also $\vert\phi_b\rangle$ by symmetry. Here, the energy required to prepare $\vert\phi_a\rangle$ is $\langle\phi_a\vert$ $H(N,\Delta,\gamma)$ $\vert\phi_a\rangle=0$ and is independent of $N,\Delta$ and $\gamma$. After this preparation, the time response is given by the solution of the $\left[ i\hbar\frac{\partial}{\partial t}-H(N,\Delta,\gamma)\right]\vert\Psi(t)\rangle=0$ time dependant \textit{Schr\"{o}dinger Wave Equation} leading after a projection on the canonical basis set used in Fig. \ref{fig:model1} to the 3 coupled equations:

 \begin{equation}  \label{eq:eff_hamiltonian1}
 i\hbar\left[\begin{matrix}
              \dot{C}_a(t)\\
              \dot{\tilde{C}}(t)\\
              \dot{C}_b(t)
              \end{matrix}
             \right]  =\left[ 
    \begin{matrix}
      0 & \sqrt{N} \gamma & 0  \\
      \sqrt{N}\gamma & \Delta & \sqrt{N}\gamma  \\
      0 & \sqrt{N} \gamma & 0
       \end{matrix}
        \right]\left[ \begin{matrix}
              C_a(t)\\
             \tilde{C}(t)\\
              C_b(t)
              \end{matrix}
              \right].
 \end{equation}
 
This system was obtained after calculating the $C_a(t), C_b(t), C_1(t),...C_N(t)$ coordinates of $\vert\Psi(t)\rangle$ on the canonical basis set, after taking into account the symmetry of the $A-N-B$ system i.e. $C_1(t)=C_2(t)=...=C_N(t)=C(t)$ and finally after performing the transformation $\tilde{C}(t)=\sqrt{N}C(t)$ as implemented in \cite{chen}.
After solving (\ref{eq:eff_hamiltonian1}) analytically, the variation in time of the $\vert\phi_b\rangle$ population amplitude is given by:
 
 \begin{equation}  \label{eq:cb1}
C_{b}(t)= N\gamma^2 \sum_{m=1} ^{3} A_m \text{e}^{i \lambda_m t/\hbar},
  \end{equation}

where $ A_m =\prod _{k \neq m} ^2   (\lambda_m - \lambda_k)^{-1}$ with $\lambda_i$ for i=1,2,3 the eigenvalues of (3). The population of the target state $\vert\phi_b\rangle$ is given by:
 
  \begin{equation} \label{eq:p1}
| C_{b}    (t) |^2 = N^2\gamma^4 \sum_{i,j=1} ^3 A_i A_j \cos \big{(}   \frac{ \lambda_i -\lambda_j }{\hbar} t  \big{)}
 \end{equation}
 
This almost periodic function leads to resonant and anti-resonant time dependant evolutions for well defined  $\gamma$ and $ \Delta$ values. For $\Delta=0$, $| C_{b} (t) |^2 $ is always periodic for all $N$. For $\Delta\neq0$ such a resonant regime is reached only when $\gamma \Delta^{-1}$ takes the values \cite{joachim87}:
  
   \begin{equation}\label{eq:pn}
 \gamma \Delta^{-1}=\left( \frac{1}{8N}\right) ^{1/2}\left[ \left(\dfrac{p}{p-2m-1} \right)^{2}-1 \right]^{1/2},  
  \end{equation}

for integer $p$ and $m$, and for $p\geqslant m+1$ and $p\neq 2m+1$. As a consequence and whatever $\gamma \Delta^{-1}$, a 1-state per line bus always permits to reach $\vert\phi_b\rangle$ from $\vert\phi_a\rangle$ in time with in average no attenuation of the $| C_{b}(t) |^2$ maximum amplitude over time as a function of $N$.

Since there is one zero eigenvalue for the reduced Hamiltonian (3), $C_{b}(t)$ is the sum of two power 2 sinusoidals of frequency $\frac{\lambda_1}{\hbar}$ and $\frac{\lambda_2}{\hbar}$. This is a generic property of a quantum bus with an odd number of states per line. Then, the $\Omega_{ab}(N)$ effective oscillation frequency between $\vert\phi_b\rangle$ and $\vert\phi_a\rangle$ is given by the largest component in (\ref{eq:p1}). For non zero $\gamma$ and $\Delta$, the largest component is $A_3 A_1$ and the secular frequency is given by: 

\begin{equation} \label{rabi}
 \Omega_{ab}(N)=-\frac{\vert \Delta \vert}{2\hbar}+\frac{1}{2\hbar}\sqrt{\Delta^2+8N\gamma^2}.
 \end{equation}
 
According to (7) and for $\gamma << \Delta$, $\Omega_{ab}(N)=2\frac{\gamma^2}{|\Delta|\hbar}N$ is linearly dependent on $N$ as already demonstrated in the spectral analysis above. For $\gamma\gg\Delta$, $\Omega_{ab}(N)$ is following a $\sqrt{N}$ moderate increase with $N$. For resonant $\Delta=0$, $\Omega_{ab}(N)=\frac{\gamma}{\hbar}\sqrt{2N}$ because here the eigenvalues of the 2 eigenstates involved in the transfer process are $\lambda_{1,2}=\pm\frac{\gamma}{\hbar}\sqrt{2N}$. Those 3 last cases were not very accessible in the above spectral analysis and are leading to an effective $V_{ab}(N)$ proportional to $\sqrt{N}$.

\end{subsection}


\begin{subsection}{ 2.2 $N$ transfer lines in parallel with 2-states per line}

On the $A-N-B$ canonical basis set $\vert\phi_a\rangle$, $\vert j \rangle$ ($j=1,...,2N$) and $\vert\phi_b\rangle$, Fig. \ref{fig:model2} is presenting the complete 2N+2 quantum states graph of the second quantum bus considered in this work with N 2-state per transfer line interacting with the emitter state $\vert\phi_a\rangle$ and the receiver state $\vert\phi_b\rangle$. Each 2-state line is $\gamma$ interacting equally with $\vert\phi_a\rangle$ and $\vert\phi_b\rangle$ and there is also a relative energy difference $\Delta$ between the quantum bus states and $\vert\phi_a\rangle$, $\vert\phi_b\rangle$. This defines three quantum $\alpha$, $\gamma$, $\Delta$ and one classical $N$ control parameters for this second A-N-B system.

    \begin{figure}[!h]
    \centering
    \includegraphics[scale=0.22]{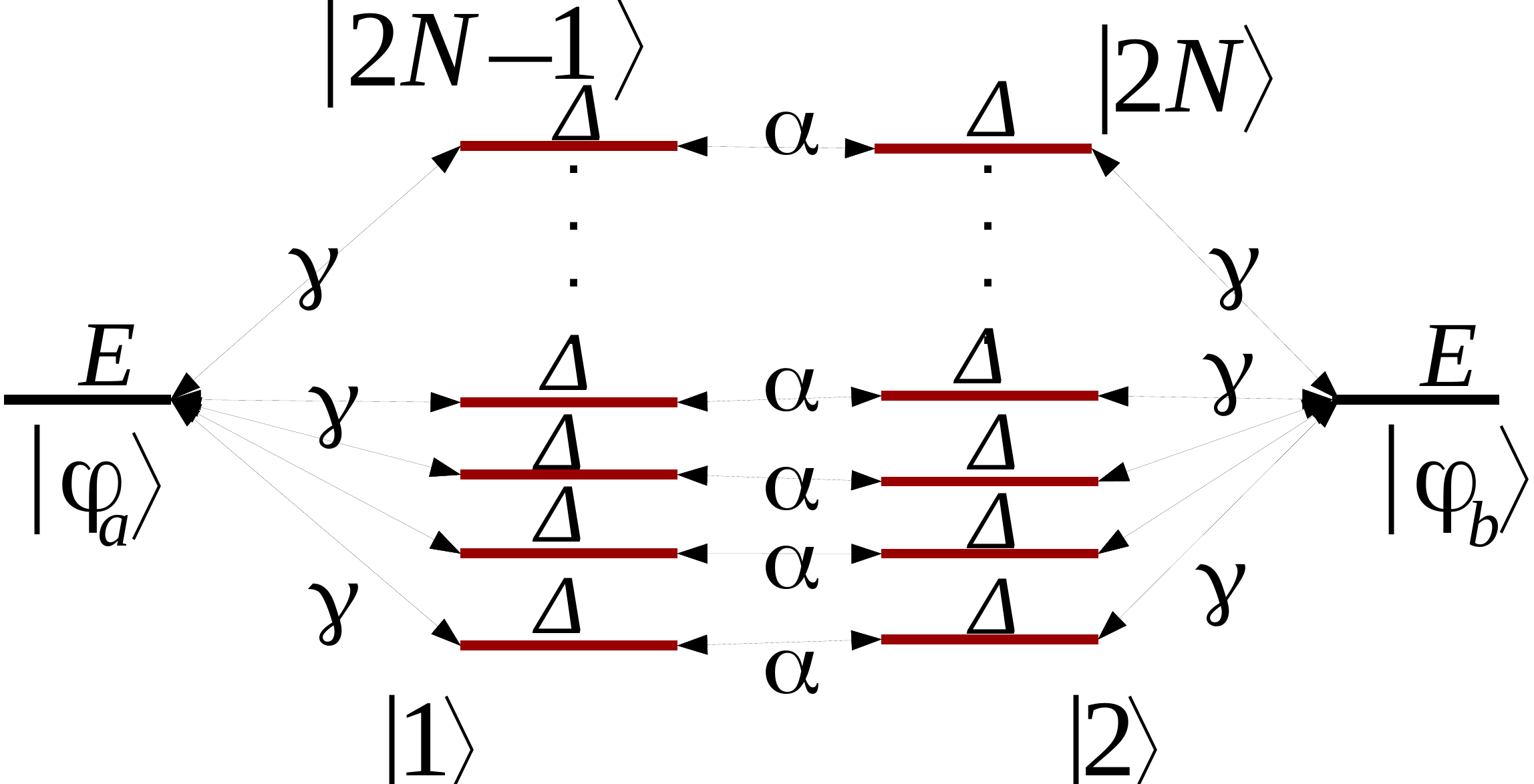} 
   \caption{The quantum graph of the $N$ 2-states per line bus interconnecting $A$ and $B$ with $\vert\phi_a\rangle$ for the electron on $A$, $\vert\phi_b\rangle$ the electron on $B$ and the $2N \vert j\rangle$ states for the electron on a given state on the bus. Those $2N$ states of the bus have the same electronic energy $\Delta$ relative to $\vert\phi_a\rangle$ and $\vert\phi_b\rangle$. They are interacting with $\vert\phi_a\rangle$ and $\vert\phi_b\rangle$ via the electronic coupling $\gamma$. $\alpha$ is the electronic coupling between 2 states along the same transfer line.}
  \label{fig:model2}
  \end{figure}
   
On its canonical basis the mono-electronic Hamiltonian of the Fig. (\ref{fig:model2}) system reads:

 \begin{equation} \label{eq:hamiltonian2_1}
H( \alpha , \gamma , \Delta , N )=\left[ \begin{matrix}
0&0&\gamma&0&\gamma&0&\cdots & \gamma&0\\
0&0&0&\gamma&0&\gamma&\cdots&0&\gamma\\
\gamma& 0 & \Delta & \alpha & 0 & 0& \cdots&0&0\\
0&\gamma&\alpha&\Delta&0&0&\cdots&0&0\\
\gamma&0&0&0&\Delta&\alpha&\cdots&0&0\\
0&\gamma&0&0&\alpha&\Delta&\cdots&0&0\\
\vdots&\vdots&\vdots&\vdots&\vdots&\vdots&\ddots&\vdots&\vdots\\
\gamma&0&0&0&0&0&\cdots&\Delta&\alpha\\
0&\gamma&0&0&0&0&\cdots&\alpha&\Delta
\end{matrix}\right] 
\end{equation}

Its spectrum has $2N+2$ eigenvalues, $N-1$ are degenerated of value $\Delta-\alpha$, $N-1$ degenerated of value $\Delta+\alpha$ and the 4 last ones are given by:

\begin{equation}  \label{spectrum3}
\begin{split}
 \lambda_{1,2}= \frac{  \Delta-\alpha \pm \sqrt{ (\Delta-\alpha)^2  +4N \gamma^2 }    }{2},\\
  \lambda_{3,4}= \frac{  \Delta+\alpha \pm \sqrt{ (\Delta+\alpha)^2  +4N \gamma^2 }    }{2}.
  \end{split}
  \end{equation}

For $\gamma << \Delta$, two cases are observed. For $\alpha\leqslant\gamma$, two of those eigenvalues, $\lambda_2$ and $\lambda_4$, have their corresponding eigenvector very close in to $\vert\phi_a\rangle$ and $\vert\phi_b\rangle$. This leads to $V_{ab}(N)\simeq \frac{\Delta N\gamma^2}{\Delta^2-\alpha^2}$, which is increasing linearly with $N$ as for a 1-state per line bus. For $\alpha>\gamma$, the two eigenstates are now the ones with their respective eigenvalues $\lambda_4$ and $\lambda_1$ leading to $V_{ab}(N)\simeq\dfrac{1}{2}\left[\Delta+\alpha+ \dfrac{\Delta N\gamma^2}{\Delta^2-\alpha^2}\right]$ also linearly depending on $N$. 
For $\gamma >> \Delta$ down to $\gamma=\Delta$, two cases are also observed. For $\alpha\leqslant\gamma$,  $V_{ab}(N)=(\lambda_{2}-\lambda_{4})/2=-\dfrac{\alpha}{2}-\dfrac{1}{4}\left[\sqrt{(\Delta-\alpha)^2+4N\gamma^2}-\sqrt{(\Delta+\alpha)^2+4N\gamma^2}\right]$. For $\alpha>\gamma$, the two eigenvalues corresponding to the two eigenvalues with their respective eigenvectors closed to $\vert\phi_a\rangle$ and $\vert\phi_b\rangle$ are now $\lambda_{1}$ and  $\lambda_{4}$  leading to $V_{ab}(N)=(\lambda_{1}-\lambda_{4})/2=-\dfrac{\alpha}{2}+\dfrac{1}{4}\left[\sqrt{(\Delta-\alpha)^2+4N\gamma^2}+\sqrt{(\Delta+\alpha)^2+4N\gamma^2}\right]$ leading finally in the 2 cases to an $\sqrt{N}$ law for $V_{ab}(N)$. This is also obtained for $\Delta=0$ leading to $V_{ab}(N)=(\lambda_{1}-\lambda_{4})/2=  -\dfrac{\alpha}{2}+\dfrac{1}{2}\sqrt{\alpha^2+4N\gamma^2}$. Finally, there are cases where this spectral analysis does not allow to determine the effective coupling $V_{ab}(N)$. For example, when $\alpha=\gamma=\Delta$, the $\lambda_4$ eigenvector has still the highest weight on $\vert\phi_a\rangle$ and $\vert\phi_a\rangle$. But at the same time,  $\lambda_1$ and  $\lambda_2$  have exactly the same weight, which makes the selection of only two eigenvectors difficult in this case.

Following protocol (ii), $\Omega_{ab}(N)$ and therefore $V_{ab}(N)$ can be determined in all cases by preparing the Fig. \ref{fig:model2} system at $t=0$ in the non stationary state $\vert\phi_a\rangle$ triggering a spontaneous response of the complete $A-N-B$ system in time. As compared to the spectral determination of $V_{ab}(N)$, the advantage of this preparation is here again that $\vert\phi_a\rangle$ is now specified and also $\vert\phi_b\rangle$ by symmetry.  After this preparation, the time response is given by the solution of the $\left[ i\hbar\frac{\partial}{\partial t}-H(N,\Delta,\gamma)\right]\vert\Psi(t)\rangle=0$ time dependant \textit{Schr\"{o}dinger Wave Equation} leading after a projection on the canonical basis set used in Fig. \ref{fig:model2} to the 4 coupled equations:

\begin{equation}  \label{eq:hamiltonian2}
 i\hbar\left[ \begin{matrix}
              \dot{C}_a(t)\\
              \dot{C}_{b}(t)\\
              \dot{\tilde{C}}_{2N-1}(t)\\
              \dot{\tilde{C}}_{2N}(t)\\
              \end{matrix}
              \right] =\left[ 
    \begin{matrix}
      0 & 0 & \sqrt{N}\gamma & 0  \\
      0 & 0 & 0 & \sqrt{N} \gamma \\
      \sqrt{N}\gamma & 0 & \Delta & \alpha\\
       0 & \sqrt{N}\gamma & \alpha & \Delta
       \end{matrix}
        \right]\left[ \begin{matrix}
              C_a(t)\\
             \ C_b(t)\\
             \tilde{ C}_{2N-1}(t)\\
             \tilde{ C}_{2N}(t)
              \end{matrix}
              \right].
 \end{equation}

Following the section 2.1 approach, this system was obtained after calculating the $C_a(t), C_b(t), C_1(t),...,C_{2N}(t)$ coordinates of $\vert\Psi(t)\rangle$ on the canonical basis set, after taking into account the symmetry of the $A-N-B$ system i.e. $C_1(t)=C_3(t)=...C_{2N-1}(t)$, $C_2(t)=C_4(t)=...C_{2N}(t)$ and finally after performing the transformation $\tilde{C}_{2N-1}=\sqrt{N}C_{2N-1}$ and $\tilde{C}_{2N}=\sqrt{N}C_{2N}$.
After solving (\ref{eq:hamiltonian2}) analytically, the variation in time of the $\vert\phi_b\rangle$ population amplitude is given for this quantum bus by:

  \begin{equation}  \label{eq:cb2}
  C_{b} (t)= N\ \alpha \gamma^2 \sum_{m=1} ^{4} B_m \text{e}^{i \lambda_m t/\hbar}
  \end{equation}
  
 where $ {\displaystyle  B_m =\prod _{k \neq m} ^3   (\lambda_m - \lambda_k)^{-1}  }$. The population of the target state is then simply given by
 
\begin{equation} \label{eq:p2}
| C_{b}     (t) |^2 = N^{2} \alpha^2 \gamma^4 \sum_{i,j=1} ^4 B_i B_j \cos \big{(}   \frac{ \lambda_i -\lambda_j }{\hbar} t  \big{)}
 \end{equation}

Contrary to the 1-state per line case, the maximum $|C_{b}(t)|^2$ population over time  in not unity for all the $N$ values. But as compared to (6), there is no analytical determination possible of the resonant and anti-resonant $\alpha$, $\gamma$ and $\Delta$ values as a function of $N$. We have not pushed further this analysis to concentrate on the dominant Heisenberg-Rabi oscillation frequency of the quantum oscillation process through this 2-states per line bus.

Since there is no zero eigenvalue in the reduced Hamiltonian (10) and since its spectrum is symmetric, $C_{b}(t)$ is the sum of 4 simple sinusoidal terms. This is a generic property of quantum bus with an even number of states per line. As a consequence, there are six different oscillation frequencies in (\ref{eq:p2}): $\Omega_{ij}=(\lambda_i-\lambda_j)/\hbar $ for $i,j=1 \cdots 4$ with $i \neq j$ weighted by $ B_i B_j$.  

For $ \alpha < \Delta $, the largest coefficient in (\ref{eq:p2}) is $B_{2}B_{4}$ with the corresponding Heisenberg-Rabi oscillation frequency $\Omega_{ab}(N)= \Omega_{24}$:

\begin{equation}  \label{eq:Rabi2}
 \Omega_{ab}(N) =\bigg{|}-\dfrac{\alpha}{\hbar}-\dfrac{1}{2\hbar} \sqrt{(\Delta-\alpha)^2+4N\gamma^2}+\sqrt{(\Delta+\alpha)^2+4N\gamma^2} \bigg{|},
 \end{equation}
 
leading for $\Delta\gg\gamma$ to $\Omega_{ab}(N)\simeq\frac{2\alpha N \gamma^{2}}{\hbar(\Delta^{2}-\alpha^{2})}$ which is linearly dependent on $N$. For $\Delta\ll\gamma$ or $\Delta=\gamma$, $\Omega_{ab}(N)$ is following a $\sqrt{N}$ moderate increase with $N$. 

For $\alpha>\Delta$, the largest coefficent in (\ref{eq:p2}) is now $B_{1}B_{4}$ with the corresponding Heisenberg-Rabi oscillation frequency $\Omega_{ab}(N)= \Omega_{14}$:

\begin{equation}  \label{eq:Rabi3}
\Omega_{ab}(N) =-\dfrac{\alpha}{\hbar}+\dfrac{1}{2\hbar}\left[\sqrt{(\Delta-\alpha)^2+4N\gamma^2}+\sqrt{(\Delta+\alpha)^2+4N\gamma^2} \right]
 \end{equation}

leading also for $\Delta\gg\gamma$ to $\Omega_{ab}(N)\simeq\frac{2\alpha N \gamma^{2}}{\hbar(\Delta^{2}-\alpha^{2})}$ which is linearly dependent on $N$. For $\Delta\ll\gamma$ or $\Delta=\gamma$, $\Omega_{ab}(N)$ is again following a $\sqrt{N}$ law. Notice that the variation of $\Omega_{ab}(N)$ as a function of $\alpha$ is not a continuous function with  an effective frequency jump for $\alpha=\Delta$. This explains the above change of the largest coefficient in (\ref{eq:p2}) between $B_{2}B_{4}$ and $B_{1}B_{4}$. 

For the resonant case $\Delta=0$, the largest coefficient in (\ref{eq:p2}) is also $B_{1}B_{4}$ leading to the corresponding $\Omega_{ab}(N)$:

 \begin{equation}  \label{eq:Rabi_4}
\Omega_{ab}(N)=\Omega_{14}=-\frac{\alpha}{\hbar}+\frac{1}{\hbar} \sqrt{\alpha^{2}+4N\gamma^{2}}
 \end{equation}
 
which is linearly dependant on $N$ for $\alpha\gg\gamma$ and is following a $\sqrt{N}$ law for $\alpha\leq\gamma$. 

Finally in the very peculiar case $\alpha=\gamma=\Delta$, the two coefficients $B_{1}B_{4}$ and $B_{2}B_{4}$ in (\ref{eq:p2}) are equals. This makes the analytical calculation of the corresponding Heisenberg-Rabi oscillation frequency very cumbersome and for $N \gg 1$, the Heisenberg-Rabi frequency becomes $\Omega_{ab}(N) = \frac{\gamma}{\hbar} \, \frac{1}{2\sqrt{N}}$.

\end{subsection}


\begin{subsection}{ 2.3 Discussion}

This above detail analysis was necessary to appreciate the richness of the time dependant quantum behaviour of 1-state and 2 states per line quantum buses. For $\gamma << \Delta$ and for the 2 types of buses, $\Omega_{ab}(N)$ and therefore $V_{ab}(N)$ is always increasing linearly with $N$. This is obtained for both the spectral (i) and the time dependent approach (ii). When $\gamma$ is approaching $\Delta$, is becoming larger or when $\Delta=0$, the spectral analysis (i) is not able to capture the richness of the large quantum mixing between $\vert\phi_a\rangle$, $\vert\phi_b\rangle$ and all the other states of the canonical basis set. In this case, the interest of preparing a non stationary initial state like $\vert\phi_a\rangle$ is to ease for the determination of $V_{ab}(N)$ via $\Omega_{ab}(N)$ i.e. when the values of the control parameters are not permitting a clear spectral identification of the eigenstate participating the most in the construction of $\vert\phi_a\rangle$ and of $\vert\phi_b\rangle$ by symmetry. Starting from $\vert\phi_a\rangle$, the determination of the Heisenberg-Rabi secular oscillation frequency is a good way to pick up over time the two pertinent eigenstates. By practicing this protocol (ii) preparation, the $\Omega_{ab}(N)$  variations with $N$ are generally showing a $\sqrt{N}$ variation which is not the initial intuitive superposition law mentioned in the introduction.

\end{subsection}

\end{section}


\section*{3) Measuring $\Omega_{ab}(N)$ using a tunnel junction}

Following now protocol (iii), the measurement of $V_{ab}(N)$ using $\Omega_{ab}(N)$ requires that $A$ and $B$ interact electronically with two metallic nano-pads $M_{A}$ and $M_{B}$ respectively using states $\vert\phi_a\rangle$ and $\vert\phi_b\rangle$ as the two pointer states of the electron transfer process from $M_{A}$ to $M_{B}$ through $A-N-B$. With no bias voltage applied to the $M_{A}$-A-N-B-$M_{B}$ junction, $M_{A}$ will sometime and randomly transfer one electron to A (or $M_{B}$ to B). In this case, no potential different results between $M_{A}$ and $M_{B}$ and the required $A-N-B$ elementary charging energy is coming from thermal fluctuations since $M_{A}$ and $M_{B}$ are necessarily in interactions with some thermal reservoirs, for example the surface supporting the $M_{A}$-A-N-B-$M_{B}$ junction \cite{ratner}. When a low bias voltage difference $V$ is applied between $M_{A}$ and $M_{B}$, a net current flows through the junction and its intensity $I(N)$ is given by the Landauer formula \cite{imry}:

 \begin{equation}  \label{eq:landauer}
I(N)= \frac{2e^2}{h} T(E_f,N)\,V
 \end{equation}
  
where $\frac{2e^2}{h}$ is the quantum of conductance.

Averaged in time, $I(N) $ results from the large number of electrons transferred events per second occurring from $M_{A}$ to $M_{A}$ through $A-N-B$ \cite{ratner}. From $A$ to $B$ through the quantum bus, each individual electron transfer event is described by an Heisenberg-Rabi time dependent quantum oscillation time as discussed in section 2. At low bias voltage, we model the quantum measurement at work on this process and performed by the $M_{A}$-A-N-B-$M_{B}$ junction by the transformation:

   \begin{equation}  \label{eq:measure_rabi}
 T(E_f,N)= \bigg{|}   \int_0 ^{\infty}    C_b(t) d\mu_h (E_f,t)       \bigg{|}^2
  \end{equation}

where $C_b(t)$ is the population amplitude of state $\vert\phi_b\rangle$ as defined in section 2 for the two bus types. With (17), the intrinsic quantum time evolution running in the junction is not eliminated but filtered and transduced to give rise to $T(E_f,N)$. 
For a low electronic coupling between $\vert\phi_a\rangle$ and $\vert\phi_b\rangle$ through the bus, different $\mu_h(E_f,t)$ transduction functions have already been proposed in the past and even a $\mu_h (E,t)$ for large V. It is generally a time dependent damping exponential to avoid any divergence when calculating (17) \cite{Ness, Sanchez, Subotnik} or to reproduce the low pass filtering effect of a tunnel junction \cite{renaud}. This is also what was anticipated by Lipmann and Schwinger \cite{lippmann} to eliminate in the model of  quantum scattering the fast time variations near and on the scattering center and to be able to work only with asymptotic states far away from this scattering center.

    \begin{figure}[!h]
    \centering
    \includegraphics[scale=0.3]{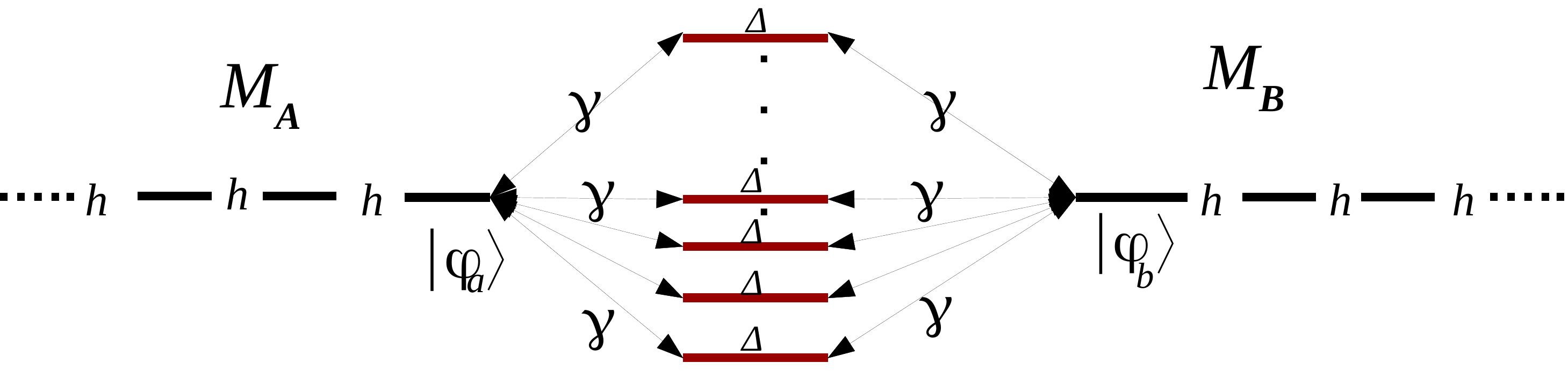} 
   \caption{ The schematic model of two semi-infinite quantum chains to model the $M_{A}$ and $M_{A}$ nano-pads interacting with the simple Fig. 1 $A-N-B$ quantum system. $h$ is the interstate coupling along those chains and $\Delta$ is the energy difference between those chains states and the central $A-N-B$ system.}
  \label{fig:measure_TB}
  \end{figure}

To determine $\mu_h (E_f,t)$, we have applied (17) to the Fig. 1 $A-N-B$ system. Here, each line of this 1-state per line bus is now interacting with 2 semi-infinite chains to model the $M_{A}$ and $M_{B}$ nano pads, each with a single conduction band  and a 4h bandwidth as presented in Fig.\ref{fig:measure_TB}. For simplicity, $\vert\phi_a\rangle$ and $\vert\phi_b\rangle$ are supposed to have the same energy than the on-site energy of an electron propagating site by site along the $M_{A}$ and $M_{B}$ chains. $C_b(t)$ and $T(E_f,N)$ can be both analytically calculated with $C_b(t)$ given by (4) and $T(E_f,N)$ given in \cite{joachim87}:

 \begin{equation}  \label{eq:t_f_1}
{T(E_f,N)}=\frac{4N^2 \gamma^4}{4 N^2 \gamma^4 + \Delta^2 h^2}.
\end{equation}

After some calculations to obtain (18) from (17) after introducing (4) in (17), the $\mu_h (E_f,t)$ measurement function reads:

  \begin{equation}  \label{eq:measure_function}
\mu_{h,\Delta}(E_f=0,t)= \frac{\sqrt{2}  h/ \hbar}{\sqrt{1-  \frac{4ih}{\Delta} }}      \text{e} ^{   -\frac{i}{2}(  \sqrt{   \Delta^2 -4i\Delta h  } -\Delta  ) t / \hbar}.
 \end{equation}
 
At $E_f$, (\ref{eq:measure_function}) can in fact be applied to any quantum system introduced in the tunnel junction if interacting only via $\vert\phi_a\rangle$ and $\vert\phi_b\rangle$ with the nano-pads. In particular when there are no quantum states in the junction, a small $\Omega$ through space electronic coupling  can remain between $M_{A}$ and $M_{B}$. This is exactly the conditions used by J. Bardeen \cite{bardeen} to get the low bias voltage V tunneling current intensity $I \propto \Omega^2 \, \rho_{  M_A   }  \, \rho_{  M_B} \, V$ through a simple $M_{A}-\Omega-M_{B}$ tunneling junction where $\rho_{  M_A   }$ and $\rho_{  M_B}$ are the $ M_A $ and $ M_B$ electronic density of states. For the simple Fig. 5 $M_A$ and $M_B$ conducting chains $ \rho_{  M_A}  = \rho_{M_B}  = \frac{1}{4 \pi h} $ leading to:

\begin{equation}  \label{eq:t1_bardeen}
 I \propto  \frac{\Omega^2}{h^2} \, V
\end{equation}

For this simple case, the corresponding $\mu_h (E_f,t)$ measurement function is given for $E_f=0$ by:

\begin{equation}  \label{eq:measure_bardeen}
d \mu_h(t)= \sqrt{2} h / \hbar \, \,\text{e}^{-ht / \hbar } dt .
\end{equation}

In this case and disconnecting now the two $M_{A}$ and $M_{B}$ measurement chains to return to the measurement protocol (ii), it remains a 2 states isolated quantum system $\vert\phi_a\rangle$ and $\vert\phi_b\rangle$ with a through space electronic coupling $\Omega$ between them. As described in section 2, preparing this simple system in the non-stationary state $\vert\phi_a\rangle$, the time variations of the $\vert\phi_b\rangle$ population amplitude during the Heisenberg-Rabi oscillation process is simply given by $C_b{(t)} = \sin {\Omega t  / \hbar}$. Then using (20) and inserting this $C_b(t)$ in (17) leads to:

\begin{equation}  \label{eq:t2_bardeen}
T(E_f)=\frac{4 \Omega^2 h^2 \hbar^2}{(h^2 +\Omega^2 \hbar^2)^2} .
\end{equation}

which is the exact $T(E_f)$ one can calculate analytically applying a simple scattering approach on a valence bond like canonical mono-electronic basis set \cite{sautet}. Interestingly, (22) reduces to (\ref{eq:t1_bardeen}) for $\Omega \hbar \ll h$ confirming that at low coupling and for this very simple $M_{A}-\Omega-M_{B}$ quantum system $T(E_f)$ is proportional to the square of the $C_b{(t)}$ oscillation frequency \cite{ratner} indicating that $\mu_h(E_f,t)$ is rather universal. Its extension for the complete energy range of the $M_{A}$ and $M_{B}$ measurement bandwidth is now under exploration.

As exemplified with (16) and also for the simple $M_{A}-\Omega-M_{B}$ Bardeen tunnel junction, (17) with (19) is able to pick up at low coupling the secular oscillation frequency of $C_b(t)$ leading to $T(E_f,N) \propto \Omega_{ab}(N)^2$. There is a limit of the functioning of this transduction because $T(E_f,N)$ is bond from above to unity and as demonstrated in section 2,   $V_{ab}(N)$ and then $\Omega_{ab}(N)$ are monotonically increasing with $N$. This limit manifests itself by the peculiar variation of $T(E_f,N)$ as a function of $\Omega_{ab}(N)$ when $\Omega_{ab}(N)$ is increasing so much that $T(E_f,N)$ is saturating to unity.  

According to (4) and (14), $C_b(t)$ is a linear superposition of sinusoidal terms with different oscillation frequencies. Since under the modulus, (17) is a linear transformation and to understand the functioning of (17), one can consider for $C_b(t)$ simply a $\sin(\Omega \, t / \hbar)$ or a $\sin^2 (\Omega t/ \hbar)$ depending respectively of the odd or even number of state in the bus lines. The unique property of (17) is that for $C_b(t)$ = $\sin(\Omega \,t / \hbar )$, $T(E_f)$  will decrease for large $\Omega$ after reaching $T(E_f)$ = 1 while for $C_b(t)=\sin^2 (\Omega t / \hbar)$, $T(E_f)$ will saturate to unity for large $\Omega$ (Figure 4). This is at the origin of the debate in the literature about the validity of the $I(N) = N^2.J$ superposition law \cite{joachim87} and \cite{lambert} since depending on the odd or even number of states per line in the quantum bus, $C_b(t)$ can be either a $\sin(\Omega \, t / \hbar)$ or a $\sin^2 (\Omega t / \hbar)$. This will be discussed in more details in the next section.

     \begin{figure}[htp]
    \centering
    \begin{tabular}{cc}
    \includegraphics[scale=0.267]{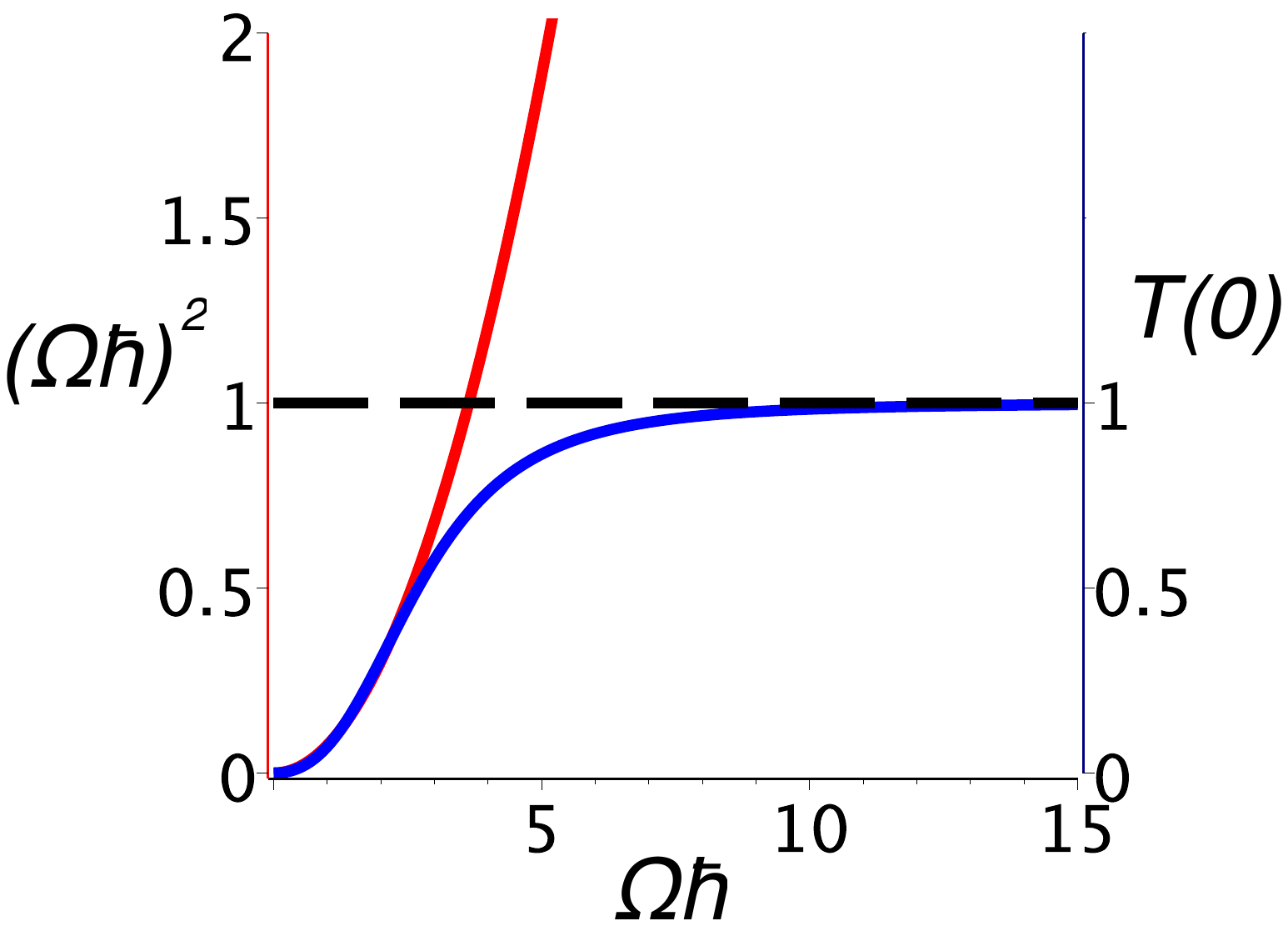} &
     \includegraphics[scale=0.25]{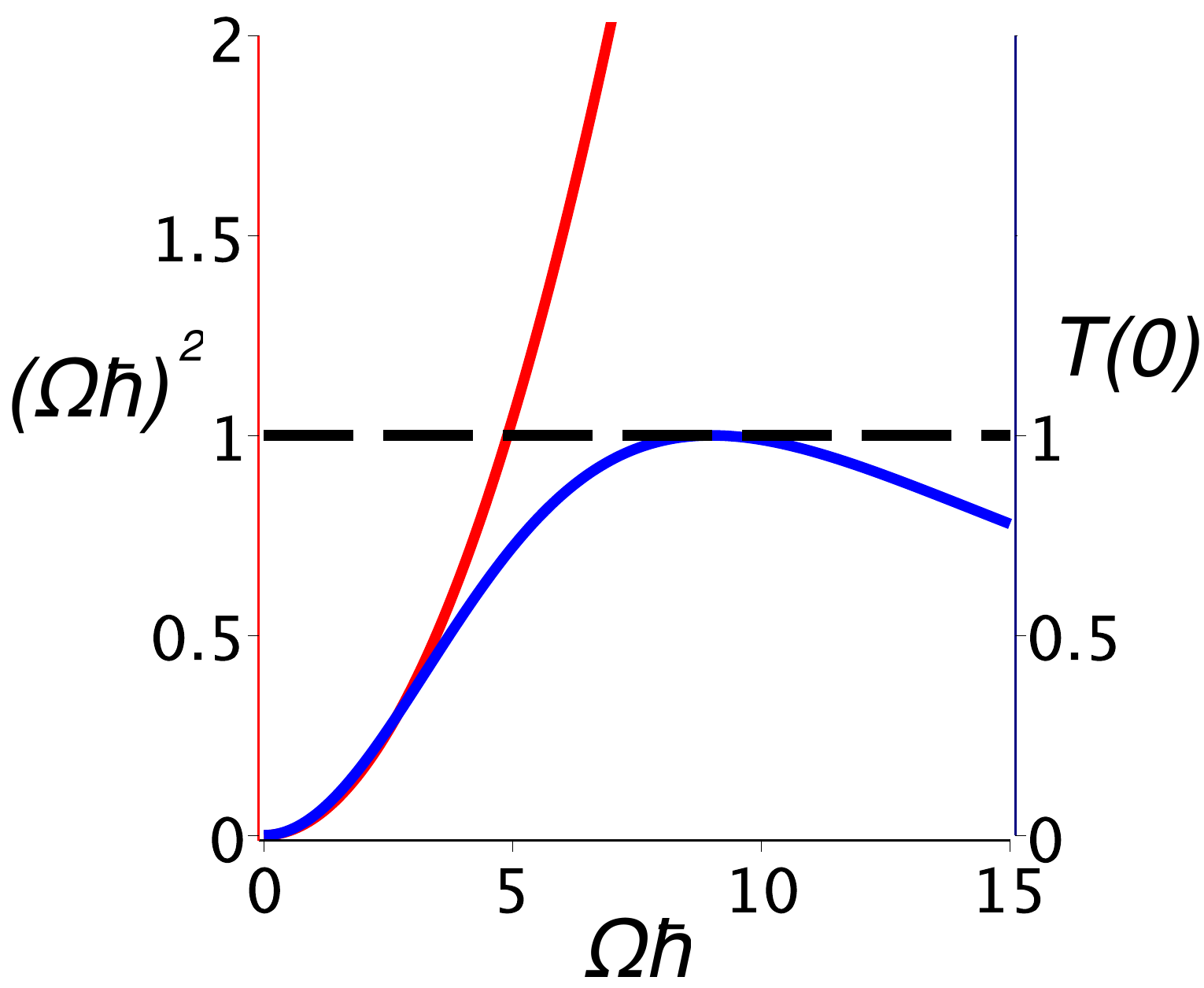} 
     \end{tabular}
   \caption{Illustration of the transduction process saturation at work in a tunneling junction. $C_b(t)$ in (17) is here simulated by $\sin(\Omega \, t / \hbar)$ for an odd number of state per line case (left) and by $\sin^2 (\Omega t / \hbar)$ for an even number of states per line (right). In both cases, the red curve is indicating the $\Omega^2$ variation expected at low coupling for $T(E_f)$. Due to the $T(E_f)$ normalisation to unity and also to the low pass filter character of the $\mu_h (E_f,t)$ transduction function in (17), $T(E_f)$ is either saturating (left) or even decaying (right) when $\Omega$ is increasing indicating the quantum limitation of this transduction process which can be tuned by changing the value of $h$ in $\mu_h (E_f,t)$.}
  \label{fig:variation2}
  \end{figure}
  
The second property of (17) is its low pass filtering character on any $C_b(t)$ due to the $\mu_h (E_f,t)$ exponential time dependant term. Already noticed in \cite{renaud}, this implies that the large $C_b(t)$ frequency components will not be capture in $T(E_f)$ because for $\Omega >  h/ \hbar$, $T(E_f)$ is first saturating to unity. Therefore, using (17) is a good way to extract the secular $C_b(t)$ oscillation frequency for a well tuned $\mu_h(E_f,t)$ function that is for a good selection of the spectral bandwidth 4h of the $M_{A}$ and $M_{B}$ nano-electrodes.


\section*{4)The parallel quantum circuit law}

Knowing the general properties of the linear transformation (17) to pass from $C_b(t)$ to its corresponding $T(E_f)$, we can now discuss how the richness of the time dependent quantum behaviours of 1-state and 2-states per line buses discussed in section 2 are preserved or not through the (17) transduction effect of protocol (iii). Starting from (4) and using (17), the $T(E_f,N)$ analytical expression is given for a 1-state per line bus by: 
 
  \begin{equation}  \label{eq:t_f_1}
{T(E_f,N)}_1=\frac{4N^2 \gamma^4}{4 N^2 \gamma^4 + \Delta^2 h^2}.
\end{equation}

and for 2-states per line buses using now (14) in (17):

    \begin{equation}  \label{eq:t_f_2}
{T(E_f,N)}_2=\frac{ 4N^2 \gamma^4 h^2 \alpha^2}{ [ N^2 \gamma^4 +(\Delta^2-\alpha^2)h^2 ]^2+4N^2 \gamma^4 h^2 \alpha^2     }.
\end{equation}

Both expressions can also be directly obtained using the Elastic Scattering Quantum Chemistry (ESQC) method
starting from a mono-electronic Hamiltonian and calculating directly the corresponding scattering matrix \cite{sautet}. By doing so, the time dependent Heisenberg-Rabi oscillations are not showing up explicitly since such scattering calculations are using asymptotic non perturbed by the central junction eigenstates of the $M_{A}$ and $M_{B}$ electrodes. This is not a problem for $\gamma << \Delta$ because in this case $\Omega_{ab}(N)$ and therefore $V_{ab}(N)$ are not large enough to saturate (17) to unity assuming that $N$ is small enough not to compensate for this small coupling through the bus. But this becomes a problem for non tunneling regime or when the number $N$ of lines in the bus is compensating for the initial low coupling through a single line. In this case, there is generally no more relation between the $T(E_f,N)$ and the generally increasing $\Omega_{ab}(N)$. In effect, for $\gamma << \Delta$ and for moderate $N$ values, (23) and (24) are leading to a $N^2$ variations for both ${T(E_f,N)}_1$ and ${T(E_f,N)}_2$ as a function of $N$ up to the point where the $N$ increase is compensating the initial small $\gamma$ value. In this case, the $N^2$ law is no more valid with a saturation of ${T(E_f,N)}_1$ whatever the large $N$ values and a decreasing of ${T(E_f,N)}_2$ for large $N$ after its saturation to unity. 

     \begin{figure}[htp]
    \centering
    \begin{tabular}{cc}
    \includegraphics[scale=0.16]{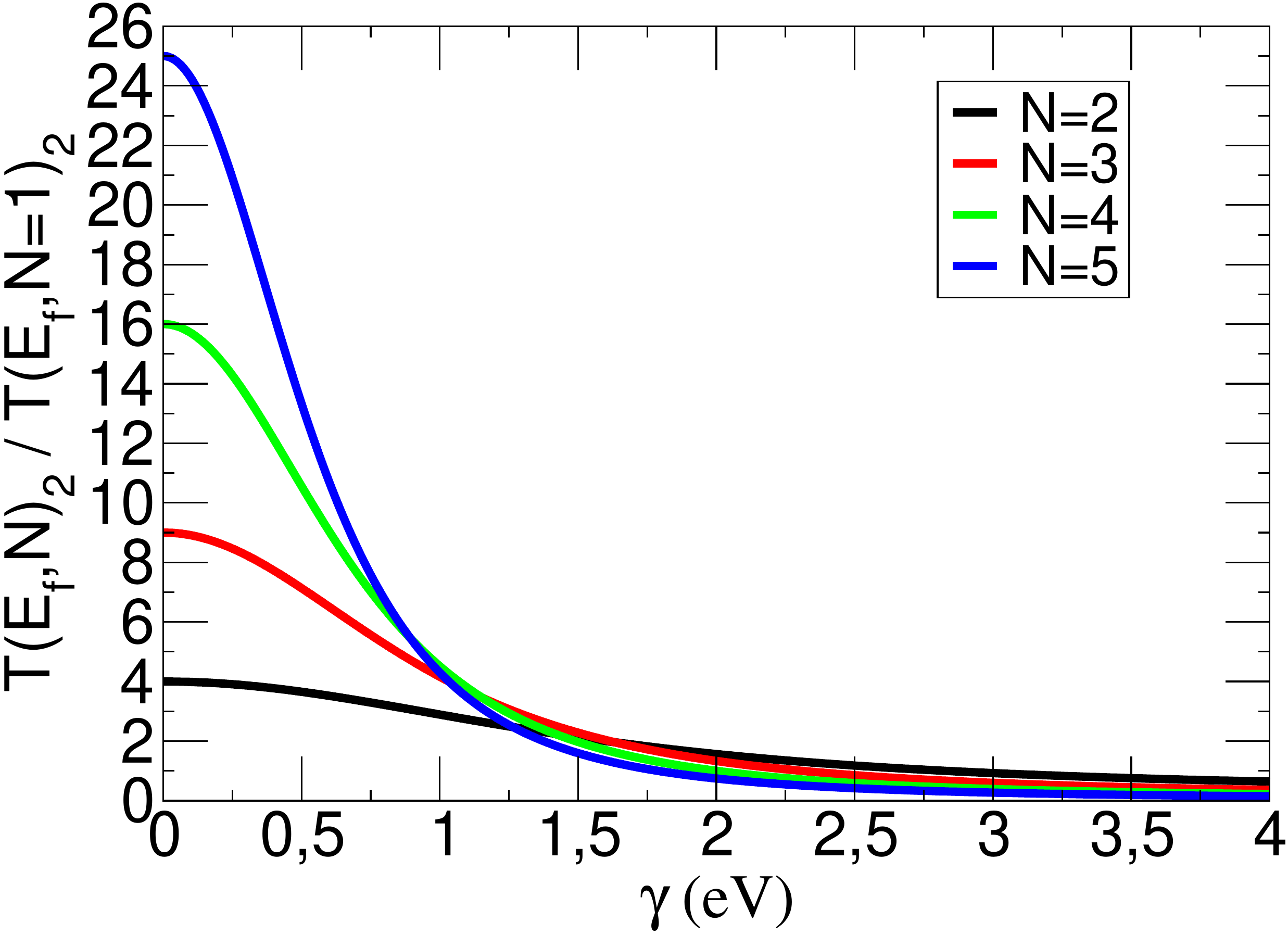} &
     \includegraphics[scale=0.16]{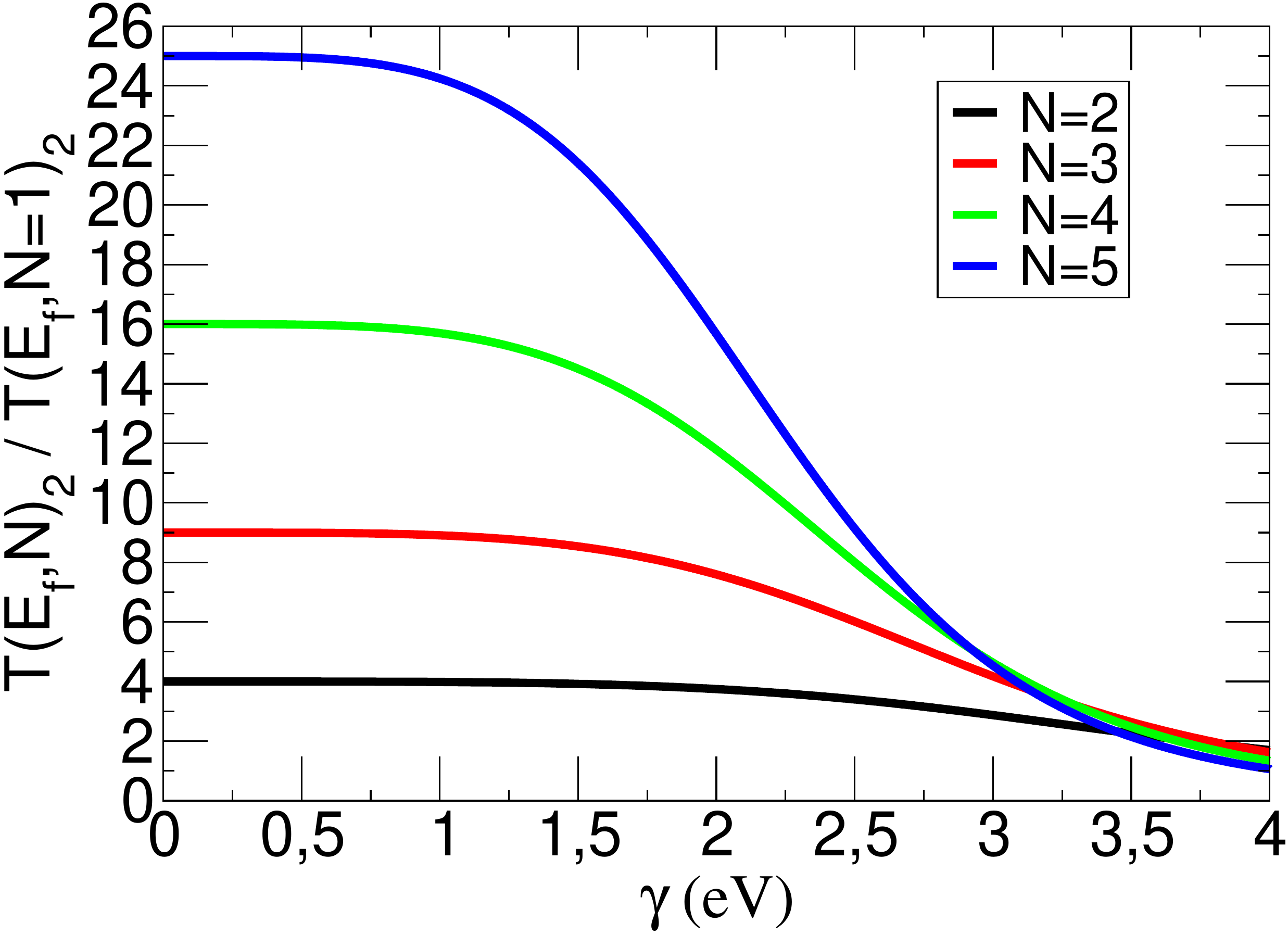} 
     \end{tabular}
   \caption{ The ratio between ${T(E_f,N)}_2$ and ${T(E_f,N=1)}_2$ for a bus with $N$ 2-states transfer lines mounted in parallel for $\Delta=0.0$ eV (left) and $\Delta=10.0$ eV (right) with $h=4.0$ eV and $N=1 \cdots 5   $ calculated at Fermi energy.}
  \label{fig:variation2}
  \end{figure}
  
${T(E_f,N)}_2$

In Figure (\ref{fig:variation2}), the range of the $N^2$ law validity is presented by plotting the $\frac{ {T(E_f,N)}_2 }{{T(E_f,N=1)}_2 }$ ratio for a 2-states perline bus. For small $\gamma$ (Figure \ref{fig:variation2} right panel), the $N^2$ law is valid at least for $N$ going from 1 to 5. But for large $\gamma$, this is no more the case. As discussed in section 3 and illustrated in Fig. 4, this is caused by the property of the transformation (17).  Notice that this $N^2$ law is valid for any odd and even number of states per line in the bus as soon as the increase in $N$ is not compensating the $\gamma << \Delta$ tunneling condition. Interestingly and also for $\gamma << \Delta$, the $N^2$ superposition law is a generalisation to N quantum conductance mounted in parallel of the $G=G_1+G_2+2\sqrt{G_1G_2}$ superposition law known in a tunnel regime for 2 quantum conductances $G_1$ and $G_2$ mounted in parallel via a quantum node \cite{magoga} since for this peculiar $N=2$ case, it comes  $ G = 4g$ for $G_1 = G_2 = g$.

More interesting are now the cases where $\gamma$ is closed or larger than $\Delta$ or when $\Delta=0$. Here, the value of h relative to $\Delta$ must also be considered because h is determining the range in energy of the (17) transduction function. Furthermore and according to (23) and (24), h is playing the same role as $\alpha$, $\gamma$ and $\Delta$ in controlling this transduction outcome. In this case and as discussed in section 2, there are many quantum control parameters values where $\Omega_{ab}(N)$ is only proportional to $\sqrt{N}$ and not to $N$. 

The case raised up by C. Lambert and co-workers and underlined in the introduction is corresponding exactly to $N = 2$ for a 2-states per line bus with $\Delta = 0$ and $\alpha = \gamma = h$ \cite{lambert}. In that case, section 2.2 is giving $\Omega_{ab}(N)  = \frac{\gamma}{\hbar}\, \frac{4N}{1+\sqrt{4N+1}} $ which is following a $\sqrt{N}$ increase with $N$. But using now (\ref{eq:t_f_2}) for this C. Lambert case, ${T(E_f,N)}_2 = ( \frac{2N}{N^2 +1} )^2$ leading for $N=2$ to ${T(E_f,N)}_2 =\frac{16}{25}$. This is a notable decrease passing from $N=1$ to $N=2$ since for $N=1$, ${T(E_f,N)}_2=1$ in this case. This clarifies the literature debate concerning the $N^2$ power law. It turns out that the case raised up in \cite{lambert} is not a tunneling case. Already for $N=2$ and since $\Delta = 0$ and $\alpha = \gamma = h$, ${T(E_f,N)}_2$ is already in its decaying regime for an $N$ increase due to the properties of the transduction function (17).

To push further the discussion using (23) and (24), it is important to notice that for 1-state per line buses, an $N^2$ term is appearing both at the numerator and denominator of (23). After the ${T(E_f,N)}_1$ saturation for large $N$, this renders difficult to follow the richness of the quantum behaviour observed in section 2 using the transduction (17) for this case, for example the $\sqrt{N}$ variations of $\Omega_{ab}(N)$ for large $N$. The $V_{ab}(N)$ variations with $N$ cannot be obtained from ${T(E_f,N)}_1$ in this case. For 2-states per line buses and aside from the C. Lambert case $\Delta = 0$, there are many other interesting Heisenberg-Rabi time-dependent quantum behavior which can be capture by (24) since there is an $N^4$ term at the ${T(E_f,N)}_2$ denominator and only $N^2$ at its numerator. In(24), the ($\Delta^2 - \alpha^2$) term is also playing a great role. For example and for the specific case $\alpha=\Delta =h=\gamma$ where $\Omega_{ab}(N) = \frac{\gamma}{\hbar} \, \frac{1}{2\sqrt{N}}$ according to section 2.2, the decreasing behaviours of $\Omega_{ab}(N)$ with $N$ is captured by (17) leading to  ${T(E_f,N)}_2 = \frac{1}{1+(N/2)^2}$. This indicates how important is the tuning of h to follow the $\Omega_{ab}(N)$ variations with $N$ i.e. to optimize the transduction process at work in a tunnel junction.


\section*{5) Conclusions}

We have started by analysing the quantum spectral properties of 1-state per line and 2-states per line buses with the objective to determine how the $V_{ab}(N)$ effective electronic coupling through such buses between an emitter and a receiver states varied as a function of the number $N$ of lines mounted in parallel to form this bus. For cases where it was spectrally difficult to determine $V_{ab}(N)$, we have re-enforced this analysis by triggering an Heisenberg-Rabi time dependent through bus quantum exchange process with an effective secular oscillation frequency $\Omega_{ab}(N)$. For this purpose, we have prepared a specific initial non-stationary pointer state and used its symmetric target pointer state to capture $\Omega_{ab}(N)$. This leads to two different $\Omega_{ab}(N)$ (and therefore $V_{ab}(N)$) regimes of variations as a function of $N$: a linear one following an intuitive superposition of electronic couplings and a $\sqrt{N}$ moderate increase as a function of $N$. In a way to substitute the initial pointer state preparation by electronically coupling the quantum bus with semi-infinite electrodes, we have discussed how the quantum transduction measurement process at work in such a tunneling junction can or not faithfully follow the variation with $N$ of the through bus $V_{ab}(N)$ effective electronic coupling. Due to the normalisation to unity of the electronic transparency of any quantum bus and to the low pass filter like character of the transduction process at work in a tunnel junction, large $V_{ab}(N)$ increase due to an $N$ increase cannot be detected by a tunneling junction. The $N^2$ superposition law is preserved for $\Omega_{ab}(N)$ (and therefore $V_{ab}(N)$) for low coupling as soon as $N$ is small enough not to compensate this small through bus coupling per line. The limitations of the quantum transduction at work in a tunneling junction is also pointing out how the broadly used concept of electrical contact between a metallic nanopad and a molecular wire may be better described in term of a quantum transduction process. This is opening the way for a better optimisation of this transduction at work in a tunneling junction playing for example with the detail band structure of the metallic nanopads in charge of this transduction to optimize the so-called contact conductance.


\section{Acknowledgements} The authors like to thank the PAMS integrated FET european project for financial support during this work.
  

\bibliographystyle{aip}

\end{document}